\documentclass[conference,compsoc]{IEEEtran}

\usepackage{hyperref}
\usepackage[capitalise]{cleveref}
\usepackage[protrusion]{microtype}
\usepackage[maxbibnames=9, minbibnames=3, backend=biber, doi=false, url=false, isbn=false, style=numeric, sorting=none]{biblatex}
\bibliography{bibliography.bib}
\usepackage{graphicx}
\usepackage{xcolor}
\usepackage{xspace}
\usepackage{enumitem}

\hypersetup{
    colorlinks,
    linkcolor={black},
    citecolor={black},
    urlcolor={blue}
}

\newcommand{\ie}{\emph{i.e.}\xspace}
\newcommand{\eg}{\emph{e.g.}\xspace}
\newcommand{\etal}{\emph{et al.}\xspace}

\newcommand{\OTH}{Technical University of Applied Sciences\xspace}
\newcommand{\SAG}{Siemens AG, Corporate Research\xspace}

\newcommand{\five}[1]{\mbox{#1-V}\xspace}
\newcommand{\riscv}{\five{RISC}}
\newcommand{\noelv}{\five{NOEL}}

\newcommand{\nop}[1]{}

\newcommand{\colfootnote}[1]{\unskip\kern-.1ex\footnote{#1}}
\newcommand{\colcomma}{\unskip\kern-.1ex,}

\newcommand{\subparagraph}[1]{\vspace{4px}\textit{#1}:\newline}

\begin{document}
\title{Static Hardware Partitioning on \riscv:\\
Shortcomings, Limitations, and Prospects}

\author{
\IEEEauthorblockN{
    Ralf~Ramsauer\colcomma\IEEEauthorrefmark{1}
    Stefan~Huber\IEEEauthorrefmark{1},
    Konrad~Schwarz\IEEEauthorrefmark{2},
    Jan~Kiszka\IEEEauthorrefmark{2} and
    Wolfgang~Mauerer\IEEEauthorrefmark{1}\IEEEauthorrefmark{2}
}

\IEEEauthorblockA{
    \IEEEauthorrefmark{1}\{ralf.ramsauer, stefan.huber, wolfgang.mauerer\}@oth-regensburg.de\\
    \OTH Regensburg, Germany
}

\IEEEauthorblockA{
    \IEEEauthorrefmark{2}\{konrad.schwarz, jan.kiszka\}@siemens.com\\
    \SAG, Munich, Germany
}
}

\maketitle

\nop
{
\the\textwidth\\
\printinunitsof{cm}\prntlen{\textwidth}\\
\printinunitsof{in}\prntlen{\textwidth}\\

\the\textheight\\
\printinunitsof{cm}\prntlen{\textheight}\\
\printinunitsof{in}\prntlen{\textheight}\\
}

\begin{abstract}
On embedded processors that are increasingly equipped with multiple
CPU cores, static hardware partitioning is an established means
of consolidating and isolating workloads onto single chips.
This architectural pattern is suitable for mixed-criticality workloads
that need to satisfy both, real-time and safety requirements,
given suitable hardware properties.

In this work, we focus on exploiting contemporary virtualisation mechanisms
to achieve freedom from interference respectively isolation between
workloads. Possibilities to achieve temporal and spatial isolation---%
while maintaining real-time capabilities---% 
include statically partitioning resources, avoiding the sharing of devices,
and ascertaining zero interventions of superordinate control structures.

This eliminates overhead due to hardware partitioning,
but implies certain hardware capabilities
that are not yet fully implemented in contemporary standard systems.
To address such hardware limitations, the customisable and configurable \riscv
instruction set architecture offers the possibility of swift, unrestricted
modifications.

We present findings on the current \riscv specification and its
implementations that necessitate interventions of superordinate control
structures. We identify numerous issues adverse to implementing our
goal of achieving zero interventions respectively zero overhead: On the design
level, and especially with regards to handling interrupts. Based on
micro-benchmark measurements, we discuss the implications of our
findings, and argue how they can provide a basis for future extensions
and improvements of the \riscv architecture.
\end{abstract}

\section{Introduction}
\label{sec:intro}
To reduce hardware costs and the overall complexity of the increasing amount of
system-on-chips (SoCs) shipped in many mass-products for industry and
consumers, consolidation of such systems onto a single chip~\cite{broy:06:icse}
is desired. This also applies to high- and mixed-criticality systems that at
least partially contain safety and real-time critical components. To ensure
safety and security, uncontrolled interference between components must be
avoided.

In general, isolation is required to guarantee non-interference between
components.  A proven means for the latter is (a) static
partitioning~\cite{martins:2020:bao, ramsauer:17:ospert, li:2019:acrn} of
resources (\ie, no sharing of devices or cores between isolated components),
and (b) zero interventions/overhead caused by the superordinate control
structure (\eg, hypervisor) during operation~\cite{ramsauer:17:ospert}.
This prevents a domain from inadvertently violating restrictions imposed by the
control structure, and thus from interfering with other critical components.

By exploiting virtualisation technologies of modern CPUs to achieve the
required isolation, we can obtain isolation guarantees directly from the
hardware.  This allows for independently running general purpose (GPOS)
and real-time (RTOS) operating systems, as well as bare-metal applications on a
SoC.

The absence of hypervisor activity is almost a necessary precondition for
freedom from interference~\cite{iso26262} with regards to real-time
capabilities and decreases the amount of hypervisor code requiring validation
or certification as needed for safety-critical environments.

During the implementation of our hypervisor
\emph{Jailhouse}~\cite{ramsauer:17:ospert}---%
which strives for zero-overhead virtualisation---%
on x86~\cite{vtx} and ARM~\cite{heiser:11:ve},
we have learned that real-world issues of hardware devices
are often underestimated.
Yet they counter our design goal of eliminating virtualisation overheads.
As safety-critical real-time systems are only a subordinate aspect in the
revenue of chip vendors, it is hard to convince them to implement required architectural changes.
On customisable architectures, such as ARM, substantial license fees may occur.

Given that the free and open \riscv architecture supports 
hardware-level virtualisation~\cite{risc-v-priv}, it is in an appropriate initial design stage~\cite{pinto:21:first}, and may be a suitable
candidate for Hardware/Software Co-Design~\cite{wolf:94:hwsw} 
activities that aim at zero-overhead virtualisation, especially 
since custom designs do not need to deal with conflicting requirements
imposed by 3rd parties in control of the architectural specification.

To identify limitations of the current \riscv hardware specifications
detrimental to our design goal, we have ported the Jailhouse hypervisor
to this emerging architecture. This sheds light on shortcomings and 
limitations of the architecture, and gives prospects on
possibly required improvements for zero-trap virtualisation.

The rest of this paper is structured as follows:
In~\cref{sec:rel}, we present related work.
Afterwards, we briefly introduce properties
of the hypervisor extensions of the \riscv architecture, and
architecture and rationale of the Jailhouse hypervisor in~\cref{sec:arch}.
In~\cref{sec:evaluation}, we describe a set of fundamental micro-benchmarks.
We quantify the impact of virtualisation overhead
on the \noelv platform, a synthesisable VHDL model of a six-core 64-bit
processor that implements the \riscv architecture, and supports hypervisor
extensions~\cite{noelv}. We discuss our findings in~\cref{sec:discussion},
and conclude in~\cref{sec:conclusion}.

\section{Related Work}
\label{sec:rel}
Before the final ratification of the hypervisor extensions for the \riscv architecture
end of 2021, Pinto \etal implemented the preliminary draft of the extension (v0.61) on
a FPGA~\cite{pinto:21:first}. As their focus is on implementing the hypervisor extensions
and customising hardware components for virtualisation, they left out time-consuming
hardware optimisations, such as optimisations for address translations.
They ported their hypervisor \emph{Bao} to \riscv to identify limitations and factors of
latency overhead, and benchmark their implementation.
As they identified interrupt handling as a major issue, they optimised hardware design
with respect to eliminating the necessity of hypervisor activity for guest interrupt
delivery.
This results in major latency decrease for IRQ handling.
The \riscv community is still in the process of researching the implications of trap-free
wired interrupt handling for virtual machines~\cite{riscv_aia}.
Performance was tested on Firesim~\cite{firesim}, a cycle-exact simulator running at 3200MHz
simulation clock, using \emph{mibench embedded benchmark suite: automotive subset}, which 
is frequently used in this area.
The benchmarks compare the performance of bare-metal vs. their enhanced hypervisor 
implementation. They further investigated performance deviations when using cache colouring
mechanisms, a technique that can be used for decreasing inter-virtual-machine interference.

In contrast to Pinto \etal, we conduct our benchmarks and measurements on real
hardware, a synthesised \riscv processor on a Xilinx Virtex UltraScale+ VCU118 FPGA.
We therefore use the final ratified specification of the hypervisor extensions.

In~\cite{caforio2021vosysmonitorv}, Caforio~\etal present VOSySmonitoRV,
a mixed-criticality solution that aims for implementing static hardware
partitioning on \riscv, by intentionally abstaining from hypervisor extension.
They justify their design decision with lack of hardware that supports virtualisation
extensions. In contrast, they exploit the machine mode privilege level of
the platform to implement means for partitioning, such as memory area isolation or
interrupt delivery.
As hypervisor extensions are ratified now, future hardware implementations
will likely optimise for ratified extensions rather than for custom exploitation of
the machine mode privilege level.

\section{Architecture}
\label{sec:arch}
\subsection{Jailhouse: Concepts and Rationale}

Safely running real-time workloads of mixed criticality on multi-core
systems~\cite{linuxlife} next to Linux is a common industrial requirement
in many domains. Contemporary multi-core platforms typically feature more 
CPU cores---Hardware Threads (HARTs) in \riscv language---than workloads, and critical tasks can be exclusively assigned to 
dedicated, isolated CPU cores.
Linux, together with its feature-rich 
ecosystem, can then execute uncritical tasks on the remaining CPU cores.

Embedded virtualisation is a promising approach for implementing safe isolation of
different workloads. Execution domains, including Linux, run as guests of a hypervisor.
This approach is, for example, implemented by
XtratuM\cite{xtratum}, NOVA~\cite{nova}, and PikeOS~\cite{pikeos}.

Static hardware partitioning is a special case of embedded virtualisation; 
it \emph{exclusively} assigns hardware resources to compute domains.
Exclusive assignment of hardware resources includes exclusive assignment of
physical CPU cores to logical domains. Hence, static hardware partitioning
assumes that available computational resources exceed required computational power.
Consequently, no scheduler is required by the hypervisor, which avoids scheduling
overhead. Virtualisation extensions ensure safe cross-domain isolation.

Our approach is based on Jailhouse, a thin Linux-based partitioning
hypervisor that targets real-world systems. Motivated by the
exokernel concept~\cite{exokernel}, our aim is to reduce the
hypervisor to a minimum level of abstraction. Our goal is to minimise
the hypervisor's interaction with guests, with the intention of
preserving key quality parameters of any guest software regardless of
if it is executed natively, or under the presence of a control structure.
With this approach, guests inherit real-time guarantees of the
underlying hardware by design.
Besides unavoidable hardware overhead due to the virtualisation of the system
(\eg, second level page table translation~\cite{drepper:costofvirtualisation}),
no further software-induced overhead due to the existence of a VMM occurs during
operation.

A small code base is a precondition for certifiability for critical environments. The reduction
of guest interaction ensures the maintenance of the platform's
real-time capabilities by design---if no interceptions take place, 
the hypervisor cannot introduce increased latencies.

Running Linux in uncritical partitions of the system is a requirement for many
real-world use cases. Therefore, we partition a booted Linux system, instead of
booting Linux on a partitioned system. This offloads complex hardware
initialisation to Linux, and ensures a small code base of the hypervisor, as
only a few platform specific drivers are required (during the operational
phase, Linux is lifted into the state of a virtual machine).

To create new isolated domains, specific hardware resources (\eg, CPUs, memory,
peripheral devices) are offlined and removed from Linux. The hypervisor is
called to create a new domain has raw access to these resources. Secondary
real-time operating systems, including Linux, or even bare-metal applications
can be loaded into the domains. Jailhouse does not paravirtualise any resources
as it exclusively assigns resources to computing domains.

The hypervisor shall only be active during its boot phase (the initialisation
of the hypervisor) and during the partitioning phase (creation, initialisation
and boot of new domains). During the operational phase (system is partitioned,
and all partitions are running), the goal is no further action by the
hypervisor.

\subsection{RISC-V Platform Virtualisation}

\subsubsection{Virtualisation Architecture}
While the \riscv platform is designed to be fully virtualisable even without
dedicated virtualisation extensions (\ie, via trap-and-emulate mechanisms), the
hypervisor extension, which allows for executing
most instructions of virtual guests natively, has recently been ratified by
\emph{RISC-V International}. \riscv implements three basic
privilege modes: (1) Machine-Mode (M-Mode) where usually the Supervisor Binary
Interface (SBI)---a BIOS-like firmware---resides, (2) Supervisor-Mode (S-Mode),
typically used for the privileged operating system and (3) User-Mode (U-Mode)
for unprivileged user-level applications. When the hypervisor extension is
active, the S-Mode is utilised by the hypervisor and called \emph{Hypervisor
extended-Supervisor (HS-Mode)}. Guests run in Virtualised Supervisor (VS-Mode),
which provides shadows of key registers to minimise interventions by
the hypervisor.

\subsubsection{Memory Management}
The memory management unit (MMU) is virtualisation aware: page tables are
resolved transparently for guests using a second translation stage for guest
physical memory to host physical memory conversion.  No hypervisor intervention
is needed for page table walks and modifications.  The two-stage address
translation process~\cite{drepper:08:acmqueue} does reduce performance,
especially on TLB misses.  However, TLB misses can be reduced by using huge
pages in the second \emph{G-Stage} translation level.  As the MMU counterpart
for IO-devices (IOMMU) is still under specification process~\cite{iommu}, its
desired memory protection features and virtualisation capabilities are lacking
for devices that use direct memory access (DMA) features. This makes direct
assignment of such devices to a guest---via techniques like interrupt
remapping---impossible, and guests cannot use such devices without 
heavy hypervisor intervention.

\subsubsection{Interrupt Controller}
An architectural weakness of current generation \riscv is the \emph{Platform
Level Interrupt Controller} (PLIC), which is the first generation standard
interrupt controller.

The \riscv hypervisor extension defines an \emph{interrupt pass-through}
mechanism. The intention is to allow interrupt requests to raise exceptions in
guests without mediation by the hypervisor. As such, this feature is
highly desirable for a hypervisor focused on the real-time domain.

Unfortunately, having been developed earlier than the hypervisor extension,
neither the \emph{Core Local Interrupt Controller} (CLINT) nor PLIC allow
for direct interrupt remapping to the guest. Any interrupt (timer,
software, external) therefore first arrives at the hypervisor (timer and
software interrupts even make another detour through M-mode), before
having to be injected into the targeted domain. Due to further design
misconceptions of the interrupt controllers, additional hypervisor
intervention is needed for the guest to mark an interrupt as being handled
(\emph{claim}) respectively handled (\emph{complete}). This is because
registers associated with claim and complete are memory mapped for multiple HARTS on
the same memory page, which means we cannot rely on the MMU for access control. Therefore a superordinate control structure is needed to
prevent a cell from (un-)intentionally interfering with any other cell's
interrupts. These problems heavily affect interrupt latency and thus real-time
capabilities.

\subsubsection{Hyperthreading}
As \riscv does not support hyperthreading, there are way less possibilities for
malicious inter-cell interference than on platforms like Intel (\eg spectre,
meltdown). However, last level caches are shared and not yet
partitionable, which opens up the possibility for influencing other cell
latencies via cache pollution~\cite{pinto:21:first}.

\section{Evaluation}
\label{sec:evaluation}
\subsection{Benchmarking Setup}

To test performance implications of the hypervisor overhead on \emph{real
hardware}, we use the \textit{Xilinx Virtex UltraScale+ VCU118 FPGA}, using the
\noelv~\cite{noelv_gaisler} bitstream, which is a synthesisable VHDL model of a
\riscv processor that implements Hypervisor Extensions. H-Extensions and the
interrupt controllers follow the final, ratified
specifications. While there is an open-source bitstream available, we used the
commercial one, which supports performance optimisations and L1 and last level
caches (LLCs). The \noelv has six HARTs, each of which has a dedicated L1
cache, while sharing a common LLC. HARTs and caches run at 100MHz.
For static hardware partitioning, we use Jailhouse as hypervisor.

We perform micro-benchmarks to quantify additional overheads resp. latencies
due to the existence of the hypervisor. All micro-benchmarks are conducted in
the following measurement scenarios:

\begin{enumerate}[label=(\Alph*)]
\item As bare-metal application without an underlying hypervisor,
\item with Jailhouse in a static partitioned execution domain (parallel to Linux),
\item As (B), but with additional load in the Linux partition.
\end{enumerate}

In scenario (A), we measure the \emph{baseline} of the raw system, that is,
overheads and latencies without the existence of a hypervisor. (A) represents
the \emph{raw noise} of the platform that we cannot fall below. Scenario (B)
represents the base overhead that exists due to the existence of the
hypervisor. Finally, scenario (C) simulates conditions in a real asymmetric
multiprocessing (AMP) environment: arbitrary load on neighbouring execution
domain to stress shared system components, such as caches or system buses.

For our micro-benchmarks, we implemented our own minimalist operating system,
which is publicly available as Open Source Software.\colfootnote{Refer to
\url{https://github.com/lfd/grinch}. We call it the \emph{Grinch}, as it
benchmarks \noelv, which---apart from being a \riscv implementation---is also
French for \emph{Christmas}.}

We selected micro-benchmarks to measure relevant code paths where the
hypervisor has to intervene active in typical real-time scenarios, such as
cyclic timer interrupts, IPIs, external interrupts and frequent firmware calls,
such as those used in \riscv for remote fences.

\begin{figure}[t]
	\includegraphics[trim={3.6cm 0 0 0 }]{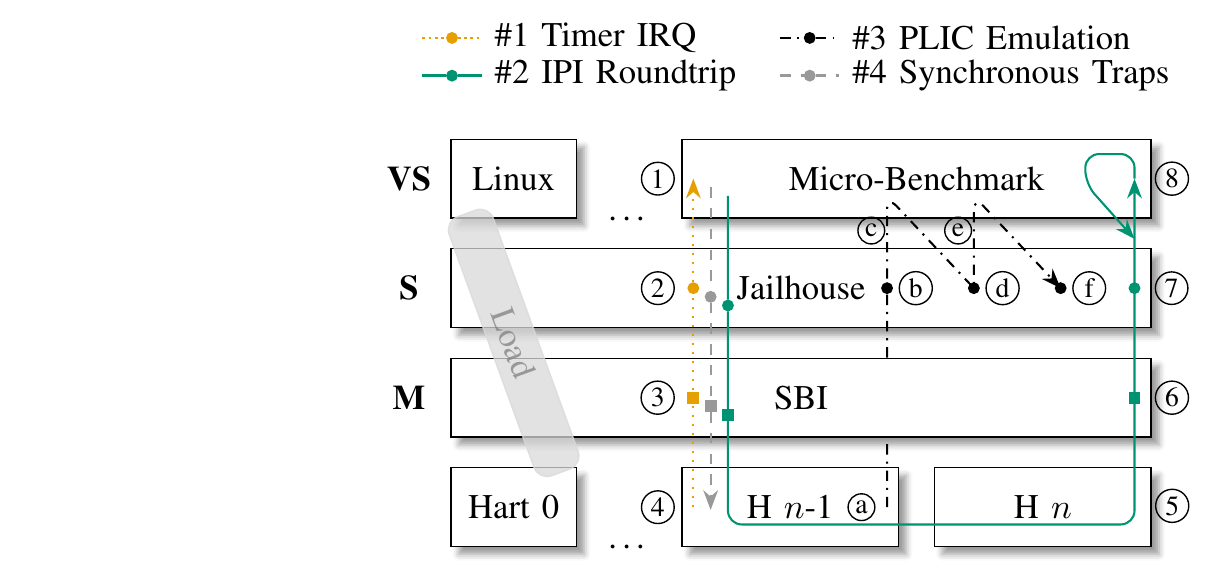}
	\caption{Illustration of the cross-systems code path for our
	benchmarks. Dots/squares mark traps; squares are unavoidable, while dots arise from HV interaction. For the IPI round trip measurement (teal),
	the path is also traversed in backward direction, as indicated by the loop.
	The additional load in the Linux domain to perturb the measurement
	is optional, and only generated in scenario (C).}\label{fig:overview}
\end{figure}

\begin{figure*}
\includegraphics{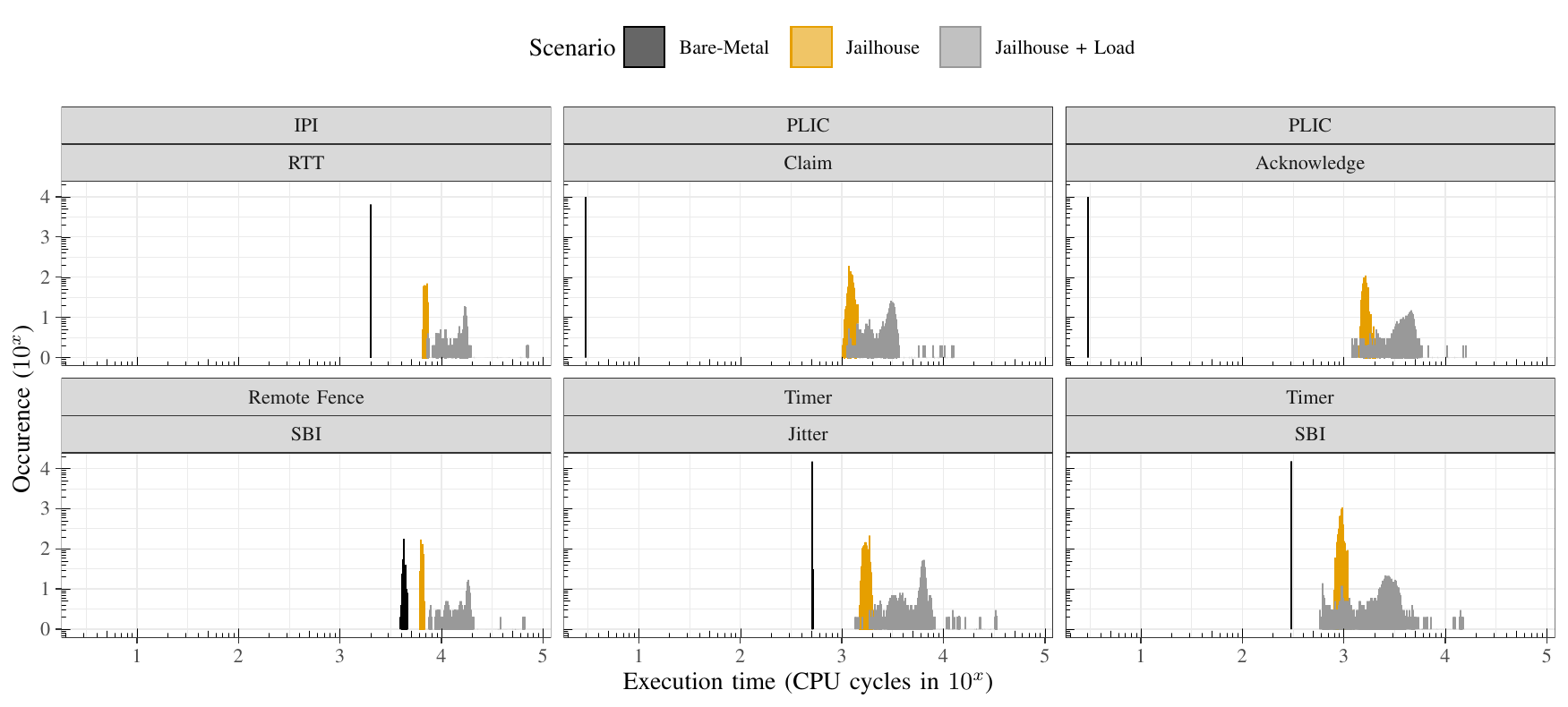}
\caption{Measurement Results (notice the double logarithmic axes): CPU cycles taken for our benchmarks---comparing performance bare-metal vs.\ with hypervisor (with and without load on other cores).}
\label{fig:results}
\end{figure*}

\subsection{Benchmark \#1---IRQ Reinjection}

As mentioned before, any IRQ on \riscv is received in S-mode and re-injected into
VS-mode. Basically, there are 3 types of IRQs: Timer, IPI, and External
interrupts (peripheral devices). External IRQs are managed by the interrupt
controller (\ie, PLIC).

In our first benchmark, we investigate timers (shown in Figure~\ref{fig:overview} in dotted ochre),
as they do not need interaction
with the PLIC, but still need to be injected by the hypervisor. Further, while
reading the current timer value can be done without hypervisor interaction,
programming the timer requires interaction with the SBI~\cite{sbispec}, which
results in moderation by the hypervisor. Any SBI call must be moderated by the hypervisor 
to ensure that the call has no cross-domain effects (\eg, CPU offlining, which
is conducted via SBI, must not affect a neighboured domain).
Typically, the overhead that is required for setting the timer only plays a
subordinate role, as the time that is required to set the timer vs. its
expiration time are significantly apart. However, for the sake of completeness,
we quantify all overheads in~\cref{fig:results}.

The essential measurement is the timer
\href{https://wiki.linuxfoundation.org/realtime/documentation/howto/tools/cyclictest/start}{jitter}: 
The difference between scheduled and actual arrival time of the timer IRQ in
(V)S-Mode. In a virtualised scenario, the hypervisor receives the timer
(2)---again via detour through the SBI (3)---and
directly injects it by setting the corresponding pending-bit (2)--(1). When the
timer arrives, our benchmark will set the next timer expiration time to a point
in future. This will automatically clear the pending flag~\cite{sbispec}.

\subsection{Benchmark \#2---IPI Round Trip Time}

As IPI Round Trip Time (RTT), shown in Figure~\ref{fig:overview} in solid teal,
we define the time that is required for sending an IPI to a secondary target
HART, and back. The only task of the target is to send the IPI back to the
initial sender. We chose this measurement, as IPIs are frequently used by
operating systems for signalling and synchronisation purposes in real-time
contexts.

On \riscv, IPIs are raised on the platform via SBI, where the target is
specified as an argument. The SBI call must be intercepted by the hypervisor
(2) as the domain membership of the target must be verified. After
verification, the IPI is propagated to the firmware (3), where it is finally
sent. From now on, the sender actively polls for the returning IPI. On receiver
side, the IPI first arrives at the hypervisor (6)---again via detour through
SBI---which injects the IPI into the guest (7)--(8) by setting the appropriate
pending bit. The guest software actively polls on the pending bit, so once the
guest sees the IPI, it sends an IPI back to the sender. The same path is
traversed backwards again: Moderation of the IPI, arrival at the sender in the
hypervisor, re-injection.

In total, four hypervisor interceptions are required for the IPI RTT
measurement: moderation for sender, arrival at receiver, moderation for
receiver, arrival at sender.

\subsection{Benchmark \#3---PLIC Emulation}

The PLIC interrupt controller offers no virtualisation possibilities.
Furthermore, the memory layout of the PLIC is unfavourably organised (\eg,
cross-hart configuration interfaces reside on the same memory page).
This requires that accesses to the PLIC must be completely emulated.

\newcommand{\rnum}[1]{\MakeUppercase{\romannumeral #1}}
The PLIC processes arriving external IRQs as follows:

\begin{itemize}
\item Physical arrival: set the external IRQ pending bit: (a)
\item Interruption of S-mode: (b)
\item \emph{Claim}ing the IRQ (\ie, read from PLIC register): (c)--(d)
\item Acknowledgement (\emph{complete}) of the IRQ (\ie, write to a PLIC
register): (e)--(f).
\end{itemize}

Under the presence of a hypervisor, the IRQ, shown in Figure~\ref{fig:overview}
in dash-dotted black, first arrives in HS-Mode. The hypervisor re-injects the
external interrupt to its guest, which will be interrupted.\colfootnote{This is the
first trap, comparable with the arrival of a timer interrupt.} The guest claims the
IRQ by reading the PLIC claim/complete register, which requires hypervisor
moderation, as well as the acknowledgement of the IRQ. The time required
for moderation can be found in~\cref{fig:results}.

\subsection{Benchmark \#4---Synchronous Traps}

Synchronous traps, shown in Figure~\ref{fig:overview} in dashed grey, arise when certain
privileged instructions are executed from less privileged modes like VS-mode. The
processor traps into higher privileged modes, where the
instructions are handled (\ie, for permission checks). We
measure the overhead of the remote fence (\emph{rfence}) firmware call, which is
frequently used to enforce ordering constraints on memory operations. It has
to detour through the SBI (3). The call from VS-Mode to SBI
is moderated by the hypervisor by trapping into S-Mode (2). We measure
synchronous traps cycle-precise using the \emph{rdcycle} instruction before and
after the trap.

\section{Discussion}
\label{sec:discussion}
In agreement with findings by Pinto \etal~\cite{pinto:21:first}, our measurements show
that real-time oriented virtualisation on \riscv---without hardware
optimisations---comes at a huge cost.  The highest cost is the unavoidable
virtualisation of the PLIC interrupt controller. Regarding complete PLIC moderation,
for example, access to the register that only requires
three cycles on bare metal may grow up to 16,000 cycles when the neighbouring
Linux domain is under load.

Another source of unnecessary overhead and complexity is the design of software
interrupts usually used for IPIs via the CLINT. Interrupt injection and
handling has to take a detour via both, S and M-Mode. This causes unnecessary
mode-changes, and thus increases latency. Additionally, IPIs cannot be distinguished:
An IPI is implemented as doorbell interrupt, and unlike with other platforms
(\eg, ARM), it is not possible to directly obtain an \emph{IPI number}.
This necessitates storing and reading the required information at a shared memory
location, which causes additional overhead.

Naturally, there are unavoidable sources of virtualisation overhead. Firstly,
cell management (\ie, creating, starting, stopping and destroying cells) 
involves hypervisor activity, yet only during system initialisation and not
during the operational phase. Secondly, additional translation for guest virtual
to guest physical addresses causes temporal overhead, even if it does not 
trigger traps.
A two-stage translation process can double TLB pressure, which
is another source of overhead. However, it can be mitigated by
using huge pages for large subsequent memory areas in the second stage.

A shortcoming of the \riscv architecture is the current generation interrupt
controller PLIC. The \riscv hypervisor extension defines an \emph{interrupt
pass-through} mechanism with the intention to allow interrupt requests to raise
exceptions in guests without intermediation by the hypervisor. As such, this
feature is highly desirable for a hypervisor focused on the real-time domain.
having been developed earlier than the hypervisor extension, the PLIC does not
support this feature.

To handle an external interrupt, three traps in the hypervisor are necessary.
This situation is supposed to be improved by the \riscv \emph{Advanced Interrupt
Architecture} (AIA)~\cite{riscv_aia}, which is currently under specification.
It addresses current limitations, and includes optimisations for virtualisation.
Both, the PLIC ant the CLINT will be superseded by advanced versions (APLIC
respectively ACLINT), and an additional interrupt controller, the \emph{Incoming
Message-Signalled Interrupt Controller} (IMSIC). One such controller
to handle MSI interrupts will be available per HART. MSIs will
be generated by writing into the IMSIC's dedicated \emph{files} for
M/S/VS-contexts.

\looseness-1 An IOMMU is currently under specification, and will allow for direct device
assignment into guests. Device MSIs can be directly (\ie, without hypervisor
intervention) forwarded to VS-mode with the IOMMU by writing into a
dedicated VS-file of the corresponding IMSIC of the target HART.

The APLIC however, will only support virtualisation partly, and wired
interrupt handling will still require trap-and-emulate mechanisms (\emph{APLIC
direct-mode}), as the \riscv community still discusses about the need and
implications of directly forwarding these kind of interrupts directly into
guests without hypervisor supervision~\cite{riscv_aia}. However, on systems
that implement both, APLIC and IMSIC(s), the APLIC can be configured to
translate wired interrupts into MSIs, which enables direct forwarding into
VS-mode in conjunction with an IMSIC (\emph{APLIC MSI-mode}).

The IMSIC can also be used as an alternative to the the (A)CLINT for sending
IPIs between virtual HARTs respectively \emph{cells}, yet without hypervisor
intervention, again by writing directly into the VS-file of the targeted
HART. Although the ACLINT will feature a separate device for S-Mode software
interrupts \emph{(SSWI)} to allow direct sending of IPIs without detour into
M-Mode, this will be still not possible for VS-mode respectively virtual
machines.

APLIC and ACLINT will not be fully optimised for
virtualisation, yet most of the problems can mostly be handled with
trickery in conjunction with the IMSIC.

\section{Conclusion}
\label{sec:conclusion}
Our work shows that static hardware partitioning on contemporary
\riscv hardware comes at a high cost. Targeted hardware optimisations for
reducing hypervisor activity have high potential for eliminating overheads
in static partitioned scenarios. Changes to the interrupt architecture (\eg,
direct interrupt handling) influence desired performance parameters more than
state-of-the-art improvements of traditionally well optimised mechanisms (\eg, MMU
2-Stage Translation). This underlines, once more, the importance of
Hardware/Software Co-Design as an important means for swift
design and implementation of useful hardware enhancements. The
\riscv ecosystem is an optimal environment for such endeavours.

Although the use-cases presented here are tailored---and crucial---to
embedded systems, the implications of our measurements are valid for
throughput-oriented general purpose systems, with significant
possible performance gains.

In future work, we will investigate the advanced interrupt architecture of the
\riscv ecosystem.

\printbibliography

\end{document}